\documentclass[12pt]{article}

\usepackage{graphics}
\usepackage{amssymb}

\newcommand{\rf}[1]{(\ref{#1})}
\newcommand{\bea}{\begin{eqnarray}}
\newcommand{\eea}{\end{eqnarray}}

\renewcommand{\l}{\lambda}

\renewcommand{\b}{\beta}

\renewcommand{\th}{\theta}
\newcommand{\oh}{\frac{1}{2}}

\newcommand{\ra}{\rangle}

\newcommand{\equ}{\!=\!}

\newcommand{\eq}{\begin{equation}}
\newcommand{\eqx}{\end{equation}}
\newcommand{\eqn}{\begin{eqnarray}}
\newcommand{\eqnx}{\end{eqnarray}}
\newcommand{\f}[2]{\frac{#1}{#2}}

\newcommand{\lm}{\lambda}

\newcommand{\Dl}{\Delta}

\newcommand{\eps}{\varepsilon}

\newcommand{\bt}{\tilde{b}}
\newcommand{\ct}{\tilde{c}}
\newcommand{\OO}{{\cal O}}
\newcommand{\VV}{{\cal V}}
\newcommand{\NN}{{\cal N}}

\newcommand{\cor}[1]{\left\langle{#1}\right\rangle}
\newcommand{\qqqq}{\quad\quad\quad\quad}
\newcommand{\ket}[1]{\left|{#1}\right\rangle}

\newcommand{\der}[2]{\frac{\partial{#1}}{\partial{#2}}}
\newcommand{\derl}[2]{{\partial{#1}}/{\partial{#2}}}

\newcommand{\qac}[1]{\left\{Q,{#1} \right\}}
\newcommand{\xb}{\bar{x}}
\newcommand{\zb}{\bar{z}}
\newcommand{\at}{\tilde{a}}
\newcommand{\dd}{{\cal D}}

\begin{document}

\begin{center}
\vspace{24pt}
{\Large \bf Interpolating between open and closed strings \\
-- a BSFT approach}

\vspace{30pt}

{\sl J. Ambj\o rn}\footnote{email: ambjorn@nbi.dk} and 
{\sl R. A. Janik}\footnote{email: janik@nbi.dk}
\footnote{Permanent address: 
M. Smoluchowski Institute of Physics, Jagellonian University, 
Reymonta~4, PL 30-059 Krakow, Poland.}

\vspace{24pt}

The Niels Bohr Institute, \\
Blegdamsvej 17, DK-2100 Copenhagen \O , Denmark\\

\vspace{48pt}

\end{center}


\begin{center}
{\bf Abstract}
\end{center}
We address the conjecture that at the tachyonic vacuum open strings
get transformed into closed strings.
We show that it is possible in the context of boundary string 
field theory to interpolate between the conventional open 
string theory, characterized by having the D25 brane as the 
boundary state, and an off-shell (open) string theory where the boundary state 
is identified with the closed string vacuum, where holomorphic 
and antiholomorphic modes decouple and where bulk vertex 
operator correlation functions are identical to  
those of the closed string.

\vspace{12pt}
\noindent


\newpage

\subsection*{Introduction}

In recent years there has been a significant progress in the
understanding  of open string tachyon condensation\footnote{For old
work on tachyon condensation, see \cite{HALP,KS}.},
i.e. the passage from the perturbative unstable open string vacuum 
to the ``true'' tachyonic vacuum with a non-vanishing tachyon
condensate\footnote{Since we consider only the bosonic string the
effective action is of course unbounded from below. What we call the   
``true'' vacuum is thus only a local minimum.}.

The tachyon condensation of the open bosonic string 
has been studied using  cubic open string field theory (OSFT) \cite{OSFT},
boundary string field theory (BSFT) \cite{BSFTW,BSFTS} and most recently
vacuum string field 
theory (VSFT) \cite{VSFT}. The use of OSFT typically involves
level truncation and quite formidable calculations. VSFT
is the conjectured form of OSFT when expanded around the tachyonic vacuum.

Of the three approaches, BSFT is on the least firm footing
as it has problems with unrenormalizable boundary interactions, but it allows 
for the most explicit verification of Sen's conjecture about 
the relation between  the 
tension of $D$-branes and the string field theory action
\cite{GERSH,KMM}.

In this paper we  discuss the conjecture that in the
tachyonic vacuum the open strings disappear and one is left with the
conventional closed string theory. There are two ways in which this can be 
realized. The simplest (and least interesting way) is the one
where the starting point is open and closed string theory in 
co-existence. 
In the process of open string tachyon condensation,
all possible excitations of the open string 
disappear and one is left with the closed string theory. 
The other possibility (which we will investigate here) is that one
starts with just the theory of  
open strings. The disappearance of the open string then 
implies the following: as one moves toward the tachyonic vacuum the 
excitations of the open string (even at tree level) should become equivalent 
to closed string excitations. 
Of course it is well known that closed strings appear in open string 
loop diagrams, but the string world sheets considered here will always
be those of open strings, in particular the above should also be true
for the disk. 

The majority of works studying the appearance of closed strings around
the tachyonic vacuum have been based on considerations of space-time
effective actions \cite{CLONE,CLTWO,CLSH}. Apart from that, 
the problem of the disappearance of open strings has been studied 
using level truncation methods in OSFT \cite{HATA,WT}, and is in a way
built into the foundation of VSFT from the start (see also \cite{CLVSFT}).
In this paper we  adopt a world-sheet perspective and use
the formalism of BSFT.

The configuration space of BSFT is the space of boundary interactions
(boundary field theories) in a certain (closed string) background,
which we here take to be flat space-time. Since a general 
boundary interaction will not result in a conformally 
invariant world-sheet theory, this is indeed a way of probing general
off-shell open strings.
Once the boundary
interactions are parameterized, there is a concrete prescription for
calculating the value of the BSFT action in terms of these parameters
(up to renormalization ambiguities).
A minimization of the BSFT action gives a flow from the unstable
perturbative vacuum to the true tachyonic vacuum.

The points in the BSFT configuration space along this flow correspond to 
two-dimensional field theories which encode how the properties of the
open string get modified in the process of decay to the tachyonic
vacuum. These theories are in general not conformal and
usually mix holomorphic and antiholomorphic fields through the
boundary interactions.

The main question that we  address in this paper is
whether in the BSFT configuration space one can find a point
which can be identified with the CFT of  ordinary
closed string theory. In particular, for such a theory there should be
a decoupling between holomorphic and antiholomorphic field modes,
contrary to the generic situation in open string theory.

We will show that one {\it can} find a continuous
interpolation between the ordinary  open string theory and an
open string theory with special boundary interactions,
where all bulk excitations (vertex operators)
are identical to closed string excitations (see below for a precise
statement), and where
holomorphic and antiholomorphic fields indeed decouple. 
Finally we perform a first analysis of the
behavior of the BSFT action for such a family of boundary interactions.

\subsection*{Setup}

Let us consider a correlation function with a number of (closed string)
on-shell vertex operators on a disk (entering the amplitude of scattering of
closed string states off an open string). For the simplest case of
(closed string) tachyons we have the well known formula
\eq
\label{e.open}
\cor{e^{ik_1 X(z_1,\zb_1)} \ldots e^{ik_n X(z_n,\zb_n)}}_{open}\sim \prod
|z_i-z_j|^{2k_i k_j} |1-z_i \zb_j|^{2k_i k_j} (1-|z_i|^2)^{k_i^2} 
\eqx 
The presence of the open string boundary can be seen through the last
two terms which mix holomorphic and antiholomorphic coordinates. 
Once we go off-shell (with respect to the open string) and add a boundary
interaction eq.\ (\ref{e.open}) is  modified.
A necessary condition for the transmutation of open string excitations 
into closed string excitations is that 
(\ref{e.open})  gets modified  to the standard closed string 
correlation function
\eq
\label{e.closed}
\cor{e^{ik_1 X(z_1,\zb_1)} \ldots e^{ik_n X(z_n,\zb_n)}}_{closed} \sim \prod
|z_i-z_j|^{2k_i k_j}
\eqx

The analysis of the tachyon condensation is usually performed
using just the quadratic tachyon
profile $T(X)=a+u X^2$ and  the end-point of tachyon condensation is for
$a\to \infty$ and $u \to 0$. At this point the amplitude
(\ref{e.open}) remains unmodified.  Additional boundary 
interactions are necessary in order to reach (\ref{e.closed}).

The goal of this paper is to construct a family of boundary
interactions which interpolates smoothly between (\ref{e.open}) and
(\ref{e.closed}).

\subsection*{Matter boundary interaction}

We consider a quadratic nonlocal interaction of the general form considered by
Li and Witten \cite{LiWitten}:
\eq
\label{e.sb}
S_B=a + \f{1}{8\pi}\int d\th d\th' X(\th) u(\th-\th') X(\th')
\eqx
where
\eq\label{u}
u(\th)=\f{1}{2\pi} \sum_{n} u_n e^{in\th}
\eqx
Our interpolating boundary interaction is defined by the choice
\eq\label{u_n}
u_n=t |n| e^{-|n|\eps}
\eqx
where $t$ is a coupling constant and $\eps$ is an UV cut-off.
In the final expressions we should set the cut-off $\eps \to 0$.

It is convenient for later use to express the boundary interaction in
terms of Fourier components of the field $X(\th)$
\eq
X(\th)=\sum_{n=-\infty}^\infty X_n e^{in \th}
\eqx
We have
\eq\label{ba}
S_B=a +\f{t}{2} \sum_{n=1}^\infty n X_{-n} X_n
\eqx
This is an `almost local' interaction. Decomposing $X=X_+ +X_- +x_0$
into positive, negative and zero modes we have a local form:
\eq
S_B=a +\f{t}{8\pi}\,i\int d\th \left(X_+ \der{X_-}{\th} -X_-
\der{X_+}{\th} \right)  
\eqx
The boundary action (\ref{e.sb}) leads to modified boundary
conditions, whose general form is presented in \cite{LiWitten}.

We will now analyze the transmutation of open strings into closed
strings through the boundary interaction (\ref{e.sb}) from three
different points of view.

\subsubsection*{I --- Green's function}

The Green's function for (\ref{e.sb}) has been derived by Li and
Witten \cite{LiWitten}.
For our choice of couplings $u_k=t \cdot |k|$, suppressing the
divergent part coming form the zero-mode, 
one is left with
\eq\label{interp}
G(z,w)=-\left(\log|z-w|^2 + \f{1-t}{1+t} \log|1-\zb w|^2 \right) 
\eqx 
It is seen that \rf{interp} interpolates between 
the propagator for the open string ($t\equ 0$) and the propagator 
for the closed string ($t \equ 1$). It follows that all correlation
functions will satisfy 
\eq
\cor{V_1(z_1)\ldots V_n(z_n)}_{\mbox{\small closed}}= \cor{V_1(z_1)\ldots
V_n(z_n)}_{\mbox{\small open with $t=1$}}
\eqx


\subsubsection*{II --- Cutting and patching holes}

Let us now consider, from a different point of view, 
the special relation between 
open strings with $t\equ 1$ boundary interactions and closed strings.

Any closed string correlation function with arbitrary
insertions {\em inside} the unit disk may be 
written as a path integral
\eq
\int \dd X \; V(z_1)\ldots V(z_n) \, e^{-S_{closed}}
\eqx
where the fields $X(z,\zb)$ are defined on the complex sphere. 
We can factorize the path integral
into an integral over the fields inside the unit disk with fixed
boundary values $X(\th)=\sum_n X_n e^{in \th}$, an integral
over the values of the fields outside the disk with the boundary values
$X(\th)$ and finally an integral over the boundary values
themselves. Let us perform first the integral outside the disk.
Following \cite{PolchBook} we split the fields outside the unit disk into a
classical piece ({\em regular} at infinity) which solves the equations
of motion and saturates the boundary conditions, and a piece
$x_D(z,\zb)$ with Dirichlet boundary conditions on the unit circle and
regular at infinity:
\eq
X(z,\zb)=X_0+\sum_{n=1}^\infty X_{-n} z^{-n}+X_n \zb^{-n} +x_D(z,\zb)
\eqx  
Substituting it into the action we get
\eq
\label{e.wave}
Normalization \cdot \exp\left(-\f{1}{2}\sum_{n=1}^\infty n X_{-n} X_n \right),
\eqx
where $Normalization$ comes from the path integral over $x_D$ (and 
is equal to the partition function for the open string with Dirichlet boundary 
conditions).  
The boundary term in eq.\ \rf{e.wave} is exactly 
our boundary action \rf{ba} with $t=1$. At
this stage we are left with an integral over the fields on the
disk with this additional boundary action, i.e.\ precisely 
an {\em  open} string correlation function with boundary 
action $S_B(t\equ 1)$.

By construction, the expression
(\ref{e.wave}) is of course just the 
Schr\"{o}dinger representation of the closed
string ground state wave function and is the simplest 
example of the operator--state mapping of conformal field
theory\footnote{In this sense one can {\it in principle} obtain
effective boundary  
interactions mimicking the insertion of arbitrary vertex operators
{\it outside} the disk, but  they will in general correspond 
to singular boundary interactions because the corresponding 
wave functions will have zeros.}.

\subsubsection*{III --- Boundary state}

We  now construct the boundary states
corresponding to our family of boundary interactions. It will enable
us to look at the meaning of the disappearance of the open string
(D25-brane) from yet another point of view. 

The prescription for associating a boundary state with a given, not
necessarily conformal, boundary interaction is \cite{YOST} 
\eq
\label{e.bs}
\ket{B}=\int dx_n d\xb_n e^{-S_b} \ket{x,\xb}
\eqx
where
\eq
\ket{x,\xb}=\prod_{n=1}^\infty e^{-\f{1}{2} |x_m|^2-a_m^\dagger
\at_m^\dagger +x_m a^\dagger_m +\xb_m \at^\dagger_m} \ket{0}
\eqx
Here $x_m=X_m/\sqrt{m}$ and $\xb_m=X_{-m}/\sqrt{-m}$.
The integration in (\ref{e.bs}) is Gaussian and the result is
\eq
\label{e.mbs}
\ket{B}=\sqrt{1+t} \, e^{-a}\prod_{n=1}^\infty \;e^{\f{1-t}{1+t} a_m^\dagger
\at_m^\dagger} \ket{0}
\eqx
The factor $\sqrt{1+t}$ comes from the infinite product $\prod_n
(1+t)^{-1}$ regularized using the $\zeta$ function prescription.

For $t\equ 0$ the boundary state can be understood as 
defining the D25  brane, or equivalently the conventional open string. 
When we increase $t$ the
properties of the D25 brane get modified as the the boundary interaction  
of open string is turned on. In particular,
the form of the boundary states shows that for $t < 1$ there 
is a coupling between holomorphic and antiholomorphic modes.
When $t$ reaches one the  holomorphic and antiholomorphic modes  decouple
and all creation operators
disappear: we are left with the closed string oscillator out-vacuum! 
Thus the interpolation between $t\equ 0$
and $t \equ 1$ may be viewed as the disappearance of the D25 brane,
and the emergence of the closed string vacuum.

\subsection*{Ghost boundary interaction}

Until now we have completely ignored the ghost sector of the 
bosonic string theory. As the matter sector is changing with 
increasing $t$ in the direction of decoupling holomorphic and
antiholomorphic modes, it is natural to assume that the ghost sector 
should also be changing. Indeed the arguments which show $S_B(t \equ 1)$ 
to originate from cutting out a disk in the closed string 
world-sheet and integrating over fields outside the disk 
can also be applied to the ghost fields. Rather than reporting
on this construction (which only applies for $t\equ 1$) we use 
the general formalism of  boundary states to construct 
a suitable ghost action which interpolates between an ordinary open
string ghost sector and one in which the holomorphic and
antiholomorphic ghosts are decoupled.

The standard ghost field mode expansions are
\eq
b(z)=\sum_n b_n z^{-n-2} \qqqq c(z)=\sum_n c_n z^{-n+1}
\eqx
and analogous ones for $\bt$ and $\ct$. In the open string the only
fields surviving on the boundary are\footnote{We write now explicitly
the tensor indices of the ghost fields, and contract them with normal
$n^a$ or tangent $t^a$ vectors to the boundary.} $b_{ab} n^a n^b$ and 
$c_t \equiv c_a t^a$ (denoted by $c(\th)$ in \cite{LiWitten}) and  
which we will
denote here by $B(\th)$ and $C(\th)/(2i)$  
respectively. These fields have the following mode expansions
\eq
B(z)=\sum_n (b_n +\bt_{-n}) e^{-in\th} \qqqq
C(z)=\sum_n (c_n -\ct_{-n}) e^{-in\th}
\eqx
The starting point of the construction of boundary states 
is to introduce classical anti-commuting (Grassmann) fields on the
boundary with Fourier components $B_n$, $C_n$ and form coherent states
(similarly to the construction in \cite{YOST}):
\eqn
\label{e.coheq}
(b_n +\bt_{-n}) \ket{B_{coh}} &=& B_n \ket{B_{coh}} \nonumber\\
(c_n -\ct_{-n}) \ket{B_{coh}} &=& C_n \ket{B_{coh}} 
\eqnx
Then the boundary state corresponding to a boundary action
$S_g(B_n,C_n)$ will be given by
\eq
\label{e.bsghdef}
\ket{B_g}=\int \prod _n dB_n dC_n \, e^{-S_g(B_n,C_n)} \ket{B_{coh}}
\eqx
where $\ket{B_{coh}}$ depends on $B_n$ and $C_n$. 

The solution of equations (\ref{e.coheq}) is
\eq
\ket{B_{coh}}=\NN \prod_{n=1}^\infty e^{-\ct_n^\dagger b_n^\dagger -C_n
b_n^\dagger -B_{-n} \ct^\dagger_n} \cdot  e^{-c_n^\dagger
\bt_n^\dagger +C_{-n} \bt_n^\dagger -B_{n} c^\dagger_n} \ket{Z}
\eqx
where $\ket{Z}=\f{c_0+\ct_0}{2}c_1\ct_1 \ket{q=0}$ (see
e.g. \cite{diVecchia}).
We fix the normalization factor $\NN$ so that the boundary state from
(\ref{e.bsghdef}) with zero action will give the ordinary open string
ghost boundary state:
\eq
\exp \left( \sum_{n=1}^\infty \left(\ct^\dagger_n
b^\dagger_n+c^\dagger_n \bt^\dagger_n \right) \right) \ket{Z}
\eqx
With this choice we have
\eq
\NN=\prod_n 4(1-\f{1}{2}C_{-n}B_n)(1+\f{1}{2}C_{n}B_{-n})
\eqx
We want now to find a ghost boundary action which will decouple the
left- and right- moving ghosts, similarly to what happened 
in the matter sector.
 
The simplest action that satisfies our requirements is
\eq
\label{e.sg}
S_g=\f{g}{2}\sum_{n=1}^\infty \left( C_{-n}B_n -C_n B_{-n} \right)
\eqx
With this choice the ghost boundary state following from
(\ref{e.bsghdef}) is
\eq
\label{e.ghbs}
\ket{B_g}=\left(\prod_n (1+g)^2 \right) \cdot 
\exp \left( \f{1-g}{1+g} \sum_{n=1}^\infty \left(\ct^\dagger_n
b^\dagger_n+c^\dagger_n \bt^\dagger_n \right) \right) \ket{Z}
\eqx
The action (\ref{e.sg}) can be rewritten as a nonlocal action of
the form
\eq
\f{1}{4\pi} \int d\th d\th' B(\th) v(\th-\th') C(\th')
\eqx
with $v(\th-\th')=\f{1}{2\pi}\sum_n v_n e^{in(\th-\th')}$ where
\eqn
v_n=-g  \qqqq n>0 \\
v_n=g \qqqq n<0
\eqnx

The above action is quite natural and involves just the combinations of ghost
fields which are present on the boundary for the ordinary open
string. The nonlocal interaction is of the same  type as
(\ref{e.sb}), and  (\ref{e.ghbs}) is the ``ghost version'' of 
(\ref{e.mbs}). However, {\em a priori}, the matter boundary
interaction with $u(\th)$ does not determine a specific ghost
interaction with some $v(\th)$.
Similarly to the situation for matter fields
we have a decoupling between holomorphic and antiholomorphic ghost
modes for $g\equ 1$.

\subsection*{Conformal properties}


It is interesting to consider the conformal properties of the boundary
interactions.  
A conformally (Virasoro-) invariant boundary state $|B\ra$ in
a boundary conformal field theory satisfies
\eq\label{cp}
L_n\ket{B}=\tilde{L}_{-n}\ket{B}
\eqx
where $L_n=\sum_k a_k a_{n-k}$. This is of course true for $t=0$. 
Our choice of matter action \rf{ba} provides a minimal 
deviation from this situation in the sense that the state
\rf{e.mbs} is still {\it SL(2,R)} invariant, 
but not Virasoro invariant
as one can 
easily check. Indeed, \rf{cp} is only satisfied for $n \equ 0,\pm 1$,
in accordance with the fact that the boundary interaction will move us
off-shell. 
The point $t \equ 1$ is special by restoring the {\it SL(2,C)} invariance 
of the closed string vacuum.

Correspondingly, if we consider the 
inclusion of ghost interactions the point $t\equ g\equ 1$
is singled out by being BRST invariant:   
\eq
Q \ket{B_{t=1}}\ket{B_{g=1}} =0
\eqx 
where $Q$ is the closed string BRST operator
\eq
Q=\sum_{n=-\infty}^\infty  c_n L_{-n}^X +\sum_{n=-1}^\infty c_{-n}
L_n^{gh} +\sum_{n=2}^\infty L^{gh}_{-n} c_n +c.c.
\eqx

\subsection*{BSFT analysis}

BSFT is defined on the space of all boundary
interactions 
\eq\label{BS1}
S_B=\sum_i \int d\th \, \lm_i \VV_i(X,b,c)
\eqx
through a choice of a corresponding family of ghost number 1
operators:
\eq\label{BS2}
\OO(\th)=\sum_i \lm_i \OO_i(\th)
\eqx
where $b_{-1}\OO_i=\VV_i$. Here $b_{-1}$ is an operator which when
expressed in terms of closed string modes is $i(b_0-\bt_0)$.
As emphasized in \cite{BSFTW} there may be an ambiguity in the choice of
$\OO_i$ for a given $\VV_i$. The BSFT action is defined as a
function of the parameters $\lm_i$ through the differential equations
\eq
\label{e.sdef}
\der{S}{\lm_i}=\f{1}{2} \int d\th \int d\th' \cor{\OO_i(\th)
\qac{\OO}(\th')} ,
\eqx
where $\cor{\ldots}$ is the {\em unnormalized} correlation function.
Since $d^2S=0$  the above equations determine, at
least locally, a well defined action. When the $\OO_i$ come from
matter weight 1 primary operators, the r.h.s. of (\ref{e.sdef}) can be
rewritten as $\b^jG_{ij}(\l)$,
where $\b^j(\l)$ are the $\b$-functions 
for the renormalization group flow in the set of boundary 
field theories, and the fixed points are defined  
by $\b^j (\l^*) \equ 0$. Our choice of coupling constants \rf{u_n}
was partially motivated by the fact that a calculation 
using the boundary interaction \rf{e.sb}-\rf{u} gives \cite{LiWitten} 
$$
\b_n(\{u_k\}) \sim \oh n(u_{n+1}-u_{n-1}) -u_n,~~~~~~~~~ n> 0.  
$$
which is zero for the choice \rf{u_n} of couplings in the limit $\eps \equ 0$.
Indeed we found earlier that the boundary state is {\it SL(2,R)}
invariant and hence scale invariant.

In principle our program is as follows: calculate the renormalization
flow (\ref{e.sdef}) 
in terms of the coupling constants $a,t,g$ and show that 
$t,g \equ 1$ is a fixed point which can be reached along a trajectory
with decreasing BSFT action. Ultimately this fixed point should have 
the following 
property: for generic boundary operators one has  
$\derl{S}{\lm_i}=0$ at $t=g=1$. This would mean that the boundary
really disappears and such perturbations could be considered as
symmetries of the theory. 

Unfortunately the calculations are non-trivial for two reasons. As noticed
already in \cite{LiWitten} there is a non-trivial UV cut-off dependence in the 
theory (here explicitly present in the innocently looking $\eps$ in 
the definition \rf{u_n} of $S_B$), as well as the need to provide an
explicit IR regularization.     The other complication is that 
the modified ghost sector should be included and it will couple in a
non-trivial way to the matter sector. The $\derl{S}{g}$ correlator is
also very complicated. While the calculation 
involving the ghosts is quite lengthy and will not 
be attempted here, let us just highlight the ambiguities mentioned 
by referring to the matter sector (we thus set $g=0$). Following
\cite{LiWitten} (except 
for a slightly different choice of ghost number 1 operator \rf{BS2} which 
is free from spurious IR divergences\footnote{We take
$\OO(\th)=\f{1}{8\pi} c_t(\th) (X(\th)-x_0)\int d\th' u(\th-\th')
(X(\th')-x_0)$ where $x_0 \equiv \f{1}{2\pi}\int d\th X(\th)$ is the
zero mode. This choice of $\OO(\th)$ ensures that the ordinary open
string with Neumann b.c. is a solution of the BSFT equations of motion
$\derl{S}{a}=\derl{S}{t}=0$ at $t=a=0$.}) one obtains from the definition
\rf{e.sdef}
\eq
\label{e.snew}
S=\left( -\sum_{m=1}^\infty  \f{\oh m(u_{m+1}-u_{m-1})-u_m}{m+u_m} -\f{1}{2}
\f{u_0+u_1}{1+u_1} +a+1 \right) \cdot Z,
\eqx  
where the partition function $Z$ is given by
\eq\label{Z}
\log Z=-\sum_{k=1}^\infty \log \left(1+\f{u_k}{k}\right) -a
\eqx
After inserting $u_k=tk e^{-k \eps}$ and isolating poles in $\eps$ we
obtain
\eq
\log Z=\f{1}{\eps} \sum_{n=1}^\infty \f{(-1)^n}{n^2} t^n
+\f{1}{2}\log(1+t) +O(\eps)-a
\eqx
We can choose to renormalize $a$ through $a=a_R+\Dl a$:
\eq
\label{e.counter}
\Dl a= \f{1}{\eps} \sum_{n=1}^\infty \f{(-1)^n}{n^2} t^n +\Dl a_{finite}
\eqx
The finite term should be fixed by some renormalization scheme
prescription (a physical definition of $a_R$). Unfortunately, as
already emphasized in \cite{LiWitten,Andreev}, we are
lacking such a prescription in the context of BSFT. We thus put 
\eq
Z=\sqrt{1+t} \cdot e^{-a_R-\Dl a_{finite}(a_R,t)}
\eqx
For $D$ scalar fields we get $\prod_{i=1}^D \sqrt{1+t_i}\cdot  e^{-a_R-
\Dl a_{finite}}$.

In an analogous way as in \cite{LiWitten} the counterterm (\ref{e.counter})
also removes the divergence in the BSFT action (\ref{e.snew}) and we 
obtain
\eq
S=\left(a_R+\Dl a_{finite}(t,a_R)+
1-\f{1}{2} \sum_{i=1}^D \f{t_i}{1+t_i}\right) \prod_{i=1}^D
\sqrt{1+t_i} e^{-a_R-\Dl a_{finite}(a_R,t)}
\eqx
From this expression it is clear that a 
well motivated  renormalization prescription for 
$\Dl a_{finite}(a_R,t)$ is needed  in order to use $S$ to study the
renormalization group flow as a function of the parameters
$a,t$. Indeed, by looking at the behaviour of the BSFT action close to
the $a=t=0$ point, one can give arguments that $\Dl a_{finite}(a_R,t)$
should be in general non-zero. 

\subsection*{Discussion}

We have tried to identify the boundary interaction $S_B$
which corresponds to the tachyonic vacuum from the 
requirement that the correlation functions in the bulk 
agree with those of the closed string. The corresponding 
(open) string theory has indeed a decoupling of 
holomorphic and antiholomorphic modes and the 
associated boundary state can be identified with 
the closed string oscillator vacuum state. All correlation functions
of bulk vertex operators coincide with those computed in an ordinary
closed string CFT.

It would be desirable to obtain this boundary interaction 
from a renormalization
group flow of the string action, i.e. to show explicitly that the BSFT
action decreases when turning on the boundary interaction. As
described above this requires
a better understanding of the regularization of BSFT and the 
corresponding renormalization conditions, as well as a complete
treatment of the ghost sector.

A better understanding of the ghost interaction and the relevant ghost
zero-mode structure 
might also throw some light on one missing step in the above construction,
namely how one obtains the closed string $S$-matrix elements 
from the open string theory with an off-shell boundary interaction. 
Although all closed string correlation functions 
of {\em any} vertex operators
are exactly reproduced with \rf{e.sb} at $t\equ 1$, we have 
of course to integrate over the positions of the vertex operators
to obtain the $S$-matrix elements and we encounter here 
a global mismatch between the integration regions of the 
closed and the open string which might be related to  
a proper definition of
the coupling of on-shell closed strings to the off-shell open string theory.
In view of the simplicity of the proposed boundary interaction, and
its very close link with the closed string CFT's 
we feel that these problems are worth further investigation.

\bigskip

\noindent{\bf Acknowledgments.}
RJ was supported in part by KBN grants 2P03B01917 and 2P03B09622. 
Both authors
acknowledge support by the
EU network on ``Discrete Random Geometry'', grant HPRN-CT-1999-00161, 
and by ESF network no.82 on ``Geometry and Disorder''.
In addition, J.A. was supported by ``MaPhySto'', 
the Center of Mathematical Physics 
and Stochastics, financed by the 
National Danish Research Foundation,

\end{document}